\documentclass[aps,prl,superscriptaddress,twocolumn,amsmath,amssymb,showpacs]{revtex4}
\usepackage{graphicx}
\usepackage{dcolumn}
\usepackage{bm}

\begin{document}

\title{Fractional time random walk subdiffusion and anomalous transport with finite mean residence times: faster, 
not slower}

\author{Igor Goychuk}
  \affiliation{Institut f\"ur Physik,
  Universit\"at Augsburg,
  Universit\"atsstr. 1,
  D-86135 Augsburg, Germany}

\date{\today}

\begin{abstract}
Continuous time random walk (CTRW) subdiffusion along with the associated fractional 
Fokker-Planck equation (FFPE) is traditionally based on the premise of random clock with 
divergent mean period. This work considers an alternative CTRW and FFPE description which is 
featured by 
finite mean residence times (MRTs) in any spatial domain of finite size. Transient subdiffusive
transport can occur on a very large time scale $\tau_c$  
which can greatly exceed mean residence time 
in any trap, $\tau_c\gg \langle \tau\rangle$, and even not being related to it. Asymptotically,
on a macroscale transport becomes 
normal for $t\gg\tau_c$. However, mesoscopic transport is anomalous. 
Differently from  viscoelastic subdiffusion no long-range 
anti-correlations among position increments are required. Moreover, our study makes it obvious that the transient
subdiffusion and transport are faster than one expects from their normal asymptotic limit on a macroscale. 
This observation has profound implications
for anomalous mesoscopic transport processes in biological cells because of macroscopic viscosity of
cytoplasm is finite.  

\end{abstract}


\maketitle

The subject of anomalous diffusion and transport became increasingly popular in the last two decades
(see Fig. 1 in \cite{G11})
with many application fields in physics and beyond and many theoretical approaches coexisting and 
developed in parallel. The 
works by Montroll and Weiss \cite{Montroll}, Scher and Montroll \cite{Scher}, and Shlesinger \cite{Shlesinger} 
laid the grounds for the continuous
time random walk  (CTRW) approach to subdiffusion based on the premise of a random distribution of the residence times
spent in traps with non-existing mean time period between two successive spatial steps. Assuming that the residence 
times in traps are independently distributed and no correlations between spatial steps exist (semi-Markovian assumption), 
the whole process
is completely characterized by a residence time distribution $\psi_i(\tau)$ in the $i$-th trap and the probability
distribution $p(k|i)$ to jump from $i$ to another trap $k$. The highly dispersive subdiffusive transport with the
position variance growing sublinearly, $\langle \delta x^2(t)\rangle \propto \kappa_{\alpha}t^\alpha$, $0<\alpha<1$,
where $\kappa_{\alpha}$ is  subdiffusion coefficient, emerges
when the residence time distributions behave asymptotically as 
\begin{eqnarray}\label{first}
\psi(\tau)\propto \tau^{-\alpha-1}.
\end{eqnarray}
This approach acquired enormous popularity and has been followed by numerous workers in the field 
\cite{Hughes,Metzler2000,Avraham}.  However, there
are no real processes in nature which would show this characteristic feature, i.e. infinite mean residence time (MRT)
in a \textit{finite} spatial domain. Of course, in reality it is always finite. Nevertheless, it can largely exceed
a characteristic time $\tau_L\sim (L^2/\kappa_{\alpha})^{1/\alpha}$ related to a finite linear size $L$ of the medium, e.g.
the width of a thin amorphous semiconductor film exhibiting akin transport properties. Then, the \textit{approximation}
of infinite mean residence time can be well justified. Asymptotically transport will be normal. However, this
macroscopic normal transport regime is irrelevant on micro and nanoscales in the corresponding amorphous media. 
The main advance of this work is to introduce an alternative \textit{semi-Markovian} CTRW approach to 
subdiffusive transport featured
but by \textit{finite} MRT as a main premise. Moreover, we derive a modified fractional Fokker-Planck equation (MFFPE) in the
continuous space limit, where MRT in a point-like trap becomes \textit{infinitely} small. Within this alternative CTRW 
approach MRT $\langle \tau\rangle$ in any finite spatial domain not only remains finite, but it shrinks to zero
with vanishing size of the domain. Nevertheless, the corresponding residence time 
distribution does
scale as one in (\ref{first}) on a long intermediate time scale. Since in all real experiments the mean residence times
are always finite one can conjecture that some experimental observations of the law (\ref{first}) along with
subdiffusion,  e.g.  in \cite{Wong},  
might be more consistent with our alternative description, rather than with the popular one.  

The key modification in this alternative is to consider the residence time distribution on sites characterized 
by the Laplace-transformed survival probability
\begin{eqnarray}\label{1}
\tilde \Phi_i(s)=\frac{1}{s+\sum_{j=0}^N r_{\alpha_j,i} s^{1-\alpha_j}},\;
\end{eqnarray} 
where $\alpha_0=1$ and $ 0< \alpha_j<1$ for $j=1,2,...,N$. By noting that $\tilde \psi_i(s)=1-s\tilde \Phi_i(s)$, 
the corresponding residence time distribution
$\psi_i(\tau)$ can be represented as a sum over $N+1$  escape channels, 
$\psi_i(\tau)=\sum_{j=0}^N\psi_{ji}(\tau)$, with the Laplace-transformed $\psi_{ji}(\tau)$ reading
\begin{eqnarray}\label{2}
\tilde \psi_{ji}(s)=r_{\alpha_j,i} s^{1-\alpha_j}\tilde \Phi_i(s).\;
\end{eqnarray} 
This justifies (see also below) to name the quantities $r_{\alpha_j,i}$ the fractional rates. One of them, 
$r_{1,i}=r_{\alpha_0,i}$ is just the normal rate corresponding to $\alpha_0=1$. The presence of such a normal escape route from
any trap is a crucial point yielding a \textit{finite} MRT,
$\langle \tau_i\rangle=\tilde \Phi_i(0)=1/r_{1,i}$ in the trap $i$. This is because the particle
can go either via any slow route, or through the fast one which defines the MRT. The MRT is thus finite in our model, but
its variance is not bounded, $\langle \delta \tau_i^2\rangle=\infty$, reflecting the presence of anomalously slow
pathways. Similar happens e.g. if the particle can escape out of the anomalous CTRW subdiffusion
domain being injected on the normal radiation boundary which it can either leave with normal rate immediately
or penetrate further inside the anomalous domain \cite{GH04}. Even more strikingly, by doing the
continuous space limit below we shall consider the limit $\langle \tau_i\rangle\to 0$. Here lies the
principal difference of our subdiffusive CTRW model from a popular one dominating currently over this research domain.
The simplest representative of this class of models
is ($N=1$)
\begin{eqnarray}\label{refined}
\tilde \Phi_i(s)=\frac{\langle \tau_i\rangle}{1+(\tau_c s)^{1-\alpha}+s\langle \tau_i\rangle},\;
\end{eqnarray} 
where $\tau_c$ is a cutoff 
 time which is equal for all traps and defines the maximal time range of subdiffusion. 
 For $\alpha=1/2$ this expression can be easily inverted to the time domain and expressed in terms of a combination of the
 error function and power law functions.
 Generally, it is not difficult to show that for $\langle \tau_i\rangle \ll \tau\ll \tau_c$, $\psi_i(\tau)$ exhibits precisely
 the same scaling behavior as one in Eq. (\ref{first}). Asymptotically for $\tau\gg \tau_c$ it is but a different power law,
 $\psi_i(\tau)\propto \tau^{3-\alpha}$ ensuring that the MRT is finite. Apart from this, initially for 
 $\tau\ll \langle \tau_i\rangle $, $\psi(\tau)\propto \tau^{-1+\alpha}$. All in all, our RTD exhibits three different
 interchanging power-laws even in the simplest case of $N=1$, providing a rather rich model.

Next, let us consider the generalized
master equation (GME) for the CTRW on the lattice $x_i=i\Delta x$ ($i\in Z$) with a discretization step $\Delta x$
and nearest neighbor jumps only, characterized by the jump probabilities $q_i^{+}=p(i+1|i)$ and 
$q_i^{-}=p(i-1|i)$ obeying $q_i^{+} + q_i^{-} =1$. The corresponding GME reads \cite{Kenkre} 
\begin{eqnarray} \label{GME}
\dot P_i(t) &=& \int_0^t \{K_{i-1}^{+}(t-t') P_{i-1}(t') +
K_{i+1}^{-}(t-t') P_{i+1}(t') \nonumber \\
&-& [K_i^{+}(t-t') + K_i^{-}(t-t')] P_i(t')\} \, \mathrm{d} t' \,
,
\end{eqnarray}
where the Laplace-transform of the kernel $K_i^{\pm}(t)$ is
related to the Laplace-transform of the residence time
distribution (RTD) by $\tilde K_i^{\pm}(s) = q_i^{\pm} \tilde
\psi_i(s) /\tilde \Phi_i(s)$, 
which for the considered model yields  
\begin{eqnarray}
\tilde K_i^{\pm}(s) = q_i^{\pm}\left ( r_{1,i}+\sum_{j=1}^N r_{\alpha_j,i}s^{1-\alpha_j} \right ),
\end{eqnarray}
where $q_i^{+}$ and $q_i^{-}$ are the forward and backward jump probabilities, accordingly,
$q_i^{+}+q_i^{-}=1$. One can define also forward an backward fractional rates so 
that $r_{\alpha_j,i}^{\pm}=q_i^{\pm}r_{\alpha_j,i}$ and $r_{\alpha_j,i}=r_{\alpha_j,i}^{+}+r_{\alpha_j,i}^{-}$,
and rewrite the GME (\ref{GME}) in the form of a fractional master equation
\begin{eqnarray} \label{FME}
\dot P_i(t) & = &\sum_{j=0}^N \sideset{_0}{_t}{\mathop{\hat
D}^{1-\alpha_j}}
 \Bigg \{r_{\alpha_j,i-1}^{+} P_{i-1}(t) \nonumber \\
&& +
r_{\alpha_j,i+1}^{-} P_{i+1}(t) 
- [r_{\alpha_j,i}^{+} + r_{\alpha_j,i}^{-}] P_i(t) \Bigg \}  \;,
\end{eqnarray}
where
\begin{eqnarray}\label{RL} 
\sideset{_{t_0}}{_t}{\mathop{\hat D}^{\gamma}}f(t):=\frac{1}{\Gamma(1-\gamma)}
\frac{\partial}{\partial t}\int_{t_0}^t dt' \frac{f(t')}{(t-t')^\gamma},
\end{eqnarray}
with $0<\gamma<1$ defines the fractional Riemann-Liouville derivative acting on some function $f(t)$ \cite{Metzler2000}.

Furthermore, let us consider an additional potential field $V(x)$ which influences the jumping rates as follows 
\begin{eqnarray}\label{form}
r_{\alpha_j,i}^{\pm}&= &(\kappa_{\alpha_j}/(\Delta x)^2)\exp [- \beta(V_{i \pm 1/2} - V_i)]
\end{eqnarray}
where 
$\kappa_{\alpha_j}$
has the meaning of a (generally) anomalous
diffusion coefficient, and $\beta=1/(k_BT)$ is inverse temperature. 
The form (\ref{form})
ensures that the Boltzmann equilibrium
condition, $r_{\alpha_j,i-1}^{+}/r_{\alpha_j,i}^{-}=\exp[\beta(V_{i-1}-V_i)]$, is satisfied for any 
$V(x)$ and our description is thermodynamically consistent. 
Using a finite difference operator $\Delta / \Delta x$,
$\Delta  P(x,t)/\Delta x :=[P(x + \Delta x/2, t) - P(x - \Delta
x/2,t)]/\Delta x$, one can rewrite the GME (\ref{GME}) for the considered model as
\begin{eqnarray} \label{FME2}
\dot P(x_i, t) &= & \left (\kappa_1+\sum_{j=1}^N\kappa_{\alpha_j}\sideset{_0}{_t}{\mathop{\hat
D}^{1-\alpha_j}} \right) \nonumber \\
&&\frac{\Delta}{\Delta x} \left ( e^{- \beta V(x_i)} \frac{\Delta}{\Delta x} \, e^{\beta
V(x_i)} P(x_i, t) \right ) \! ,
\end{eqnarray}
where we have explicitely singled out the normal transition channel which is present between any two localized states 
and acting in parallel to all other ultraslow channels.

In the continuous space
limit $\Delta x\to 0$ with constant $\kappa_{\alpha_j}$ Eq. (\ref{FME2}) yields the
 following generalization of 
the fractional Fokker-Planck equation (FFPE) of Refs. \cite{FFPE,Metzler2000,Barkai2000,G06}
\begin{eqnarray}\label{FFPE}
\frac{\partial P(x,t)}{\partial t}& = & \left (\kappa_1+\sum_{j=1}^N\kappa_{\alpha_{j}}\sideset{_0}{_t}{\mathop{\hat
D}^{1-\alpha_j}} \right )\nonumber \\
&& \frac{\partial}{\partial x}\left (e^{-\beta V(x)}
\frac{\partial}{\partial x}e^{\beta V(x)} P(x,t)\right)\;.
\end{eqnarray}
The generalization consists primarily in the presence also of a normal diffusion term along with the anomalous 
diffusion contributions. Notice that it was not only obtained from the CTRW possessing finite MRT in
each trap but even in the limit $\langle \tau_i\rangle \propto (\Delta x)^2/\kappa_1\to 0$, so that the MRT in 
any spatial domain of
a finite size is obviously finite. For $N=1$, and in the limit $\kappa_1\to 0$ it reduces to the original FFPE of works 
\cite{FFPE,Barkai2000}. In the same limit, but for $N=2$ it reduces to the modified FFPE of Ref. 
\cite{SokolovKlafter}.  Generally, it belongs to the class of distributed
order fractional kinetic equations \cite{SokolovKlafter,Chechkin}.

In the following, 
we consider the simplest case of $N=1$, with finite $\kappa_1$ and 
$\kappa_{\alpha}=\kappa_{1}\tau_c^{1-\alpha}$ with a very large $\tau_c$.
The key point is that for $t\ll \tau_c$ the normal diffusion term is negligible
and the dynamics is subdiffusive, whereas the mean residence time in any finite spatial domain is finite.
Here lies a profound, striking difference with the earlier approach. For $t\gg \tau_c$, the normal diffusion
contribution dominates the transport, which becomes asymptotically normal. However,
the biased subdiffusion changes asymptotically into superdiffusion (see below) because the variance of residence times
is infinite. 

\begin{figure}[t]
\centering
\includegraphics[width=7.5cm]{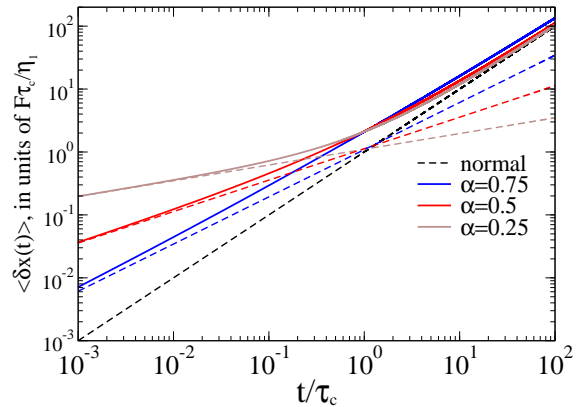}
\caption{(Color online) Anomalous transport which asymptotically becomes normal.
} 
\label{Fig2}
\end{figure}

Our focus is now on an intermediate subdiffusive regime.
Considering the mean displacement under the constant force $F$, the solution
of (\ref{FFPE}) yields
\begin{eqnarray}\label{solution1}
\langle \delta x(t)\rangle &= &Ft/\eta_1+Ft^{\alpha}/[\eta_{\alpha}\Gamma(1+\alpha)]\nonumber \\
&= &\frac{F\tau_c}{\eta_1}
\left [\frac{t}{\tau_c}+ \frac{1}{\Gamma(1+\alpha)}\left (\frac{t}{\tau_c}\right )^{\alpha}\right ] \;,
\end{eqnarray}
where the normal, $\eta_1$, and fractional, $\eta_\alpha$, frictional coefficients are related to the 
corresponding diffusion
coefficients by standard, $\eta_{1}=k_BT/\kappa_{1}$, and generalized, $\eta_{\alpha}=k_BT/\kappa_{\alpha}$,
Einstein relations, respectively.
Fig. \ref{Fig2} displays this solution for three different values of $\alpha$ along with one and the same normal diffusion
asymptotics for $t\gg \tau_c$ and three different initial $t\ll \tau_c$ asymptotics.  One can conclude that the motion remains
subdiffusive for most time and can be characterized by a time-dependent power-exponent $\alpha_{\rm eff}(t)$
which gradually changes from $\alpha$ for $t\ll\tau_c$ to $\alpha=1$ for $t\gg\tau_c$. Even more strikingly, in the 
subdiffusive regime the transport is as a matter of fact faster than its normal diffusion asymptotics. Indeed, 
already the structure of the generalized master equation (\ref{GME}) says that in the Markovian approximation, 
$P_i(t')\approx P_i(t)$, only
the normal rate term contributes asymptotically, $t\to\infty$. This is because the terms in the integral of the memory kernel, 
$\int_{0}^t K_{i}^{\pm}(t')dt'$, which 
correspond to the anomalous rate contributions, tend to zero with increasing the upper limit $t$, 
as $1/t^{1-\alpha_j}$, 
being always 
positive. Initially these terms, however, profoundly contribute making the overall transport faster than 
its normal asymptotics.  Notice that this interpretation is opposite to one suggested in Ref. \cite{Chechkin}
for diffusion which we consider shortly: the transport obviously slows down in the course of time and not
accelerates.

For the displacement variance $\langle \delta x^2(t)\rangle =\langle x^2(t)\rangle-
\langle \delta x(t)\rangle^2$,  which characterizes diffusion, after some algebra we obtain
\begin{eqnarray}\label{solution2}
\langle \delta x^2(t)\rangle &= &  2\kappa_1 t+2\kappa_\alpha t^{\alpha}/\Gamma(1+\alpha)\\
&& +\left( \frac{F\tau_c}{\eta_1}\right)^2
\Bigg \{ \frac{2(1-\alpha)}{\Gamma(2+\alpha)}\left(\frac{t}{\tau_c}\right)^{1+\alpha}  \nonumber \\
&+ & \left [ \frac{2}{\Gamma(1+2\alpha)}-\frac{1}{\Gamma^2(1+\alpha)} \right] \left (\frac{t}{\tau_c}\right )^{2\alpha}
 \Bigg \} \;. \nonumber
\end{eqnarray}
Taking into account the generalized Einstein relation, 
it is seen immediately that a linear fluctuation-response relation, 
$\langle \delta x(t)\rangle=[F/(2 k_BT)]\langle \delta x^2(t)\rangle_{F=0}$ is satisfied, between the 
``fluctuation'' $\langle \delta x^2(t)\rangle_{F=0}$ in the absence of driving force $F$, and the ``response''
$\langle \delta x(t)\rangle$ to it. In the subdiffusive transport regime restricted by $t\ll\tau_c$, 
\begin{eqnarray}\label{approx}
\langle \delta x(t)\rangle &\approx & F t^{\alpha}/[\eta_{\alpha}\Gamma(1+\alpha)], \nonumber \\
\langle \delta x^2(t)\rangle &\approx & 2\kappa_\alpha t^{\alpha}/\Gamma(1+\alpha),
\end{eqnarray}
i.e. the variance $\langle \delta x^2(t)\rangle$ behaves subdiffusively also in the presence of bias $F$, like
the mean displacement in Fig. \ref{Fig2}.
This is in a striking contrast with the standard CTRW and FFPE subdiffusion based on the premise
of infinite mean residence times. In the latter case,  $\langle \delta x^2(t)\rangle_{F\neq 0}\propto F t^{2\alpha}$,
for a sufficiently large $F$, or asymptotically,
similar to the last term in Eq. (\ref{solution2}). Therefore, our theory is closer to what
one intuitively expects for normal transport, sharing this normal feature also with the 
alternative Generalized Langevin Equation (GLE) approach to subdiffusion which we shall discuss shortly.

 Within the presented
modified theory the large time asymptotic behavior of the variance in the presence of bias is also rather unusual.
Obviously for $F\neq 0$, and $t\gg \tau_c$, the main contribution in Eq. (\ref{solution2}) is superdiffusive,
$\langle \delta x^2(t)\rangle_{F\neq 0}\propto F t^{1+\alpha}$. The explanation of this is the following.
Even if the random clock governing subdiffusion in our case does possess a mean value, its variance is but
infinite and in accordance with Ref. \cite{G11} (see Appendix therein) determines the displacement variance 
in the case of \textit{biased} diffusion. Unbiased diffusion remains but asymptotically normal. 
Such a combination of normal transport and superdiffusion was found recently 
in numerical simulations of normal Langevin dynamics in near critically biased washboard potentials in the presence of frozen
disorder for some parameters \cite{Khoury}. In this respect it is worth mentioning that hopping diffusion in disordered 
potentials is equivalent
within the mean-field (effective medium) approximation to a CTRW \cite{Hughes}. This provides one of physical
justifications for the CTRW phenomenological approach.

Let us compare now the considered modification of CTRW subdiffusional transport phenomenology with even more normal route of
the GLE approach to subdiffusion \cite{Kubo,Zwanzig,G09,G12},
where all the moments of the residence time distribution in any finite-size spatial domain are finite \cite{G09}.
Such a GLE approach is intrinsically based on the phenomenon of viscoelasticity \cite{Maxwell,Gemant} 
with long-range
correlations in the Brownian particle displacements. Here, the physical origin of subdiffusion is entirely different 
from one in the semi-Markovian CTRW \cite{G09,G12}. Stochastic dynamics is described by the overdamped GLE \cite{Kubo,Zwanzig} 
 \begin{eqnarray}\label{GLE}
\int_0^t\eta(t-t')\dot x(t')dt'=-\frac{\partial V(x,t)}{\partial x}+\xi(t),
\end{eqnarray}
where thermal force $\xi(t)$ is (i) Gaussian, (ii) unbiased on average, and (iii) obeys
 the fluctuation-dissipation relation,
\begin{eqnarray}\label{FDR}
\langle \xi(t)\xi(t')\rangle = k_B T \eta(|t-t'|),
\end{eqnarray}
at the environmental temperature $T$. Macroscopic power-law models of viscoelasticity were introduced in 
1936 by A. Gemant \cite{Gemant,G07} as a generalization of the Maxwell model \cite{Maxwell} with exponentially decaying
viscoelastic memory kernel. One of the Gemant's macroscopic models having relation to
our present work corresponds to the memory friction
\begin{eqnarray}\label{Gemant1}
\tilde \eta(s)=\frac{\eta_1}{1+(s\tau_c)^{1-\alpha}},
\end{eqnarray}  
in the Laplace space. For $t\ll \tau_c$,  $\eta(t)$ can be approximated by 
$\eta(t)\approx \eta_{\alpha}/[\Gamma(1-\alpha)t^\alpha]$, with $\eta_{\alpha}=\eta_1\tau_c^{\alpha-1}$ and 
its asymptotic decay for $t\gg \tau_c$ 
is another power law, $\eta(t)\propto t^{\alpha-2}$ ensuring that $\tilde \eta(0)=\eta_1$ is finite. 
This GLE model with the memory kernel (\ref{Gemant1}) for $f(x)=F=const$ yields
again exactly the same solution (\ref{solution1}) for the mean displacement and a modified (with $F\to 0$,
the first line only remains) solution
(\ref{solution2}) for the variance  \cite{FNL12}.
It must be emphasized that GFFPE (\ref{FFPE}) with $N=1$ does not present a master equation counterpart of the GLE (\ref{GLE})
with the memory kernel (\ref{Gemant1}), as they have physically (and mathematically too)  totally different origins.
The both approaches share, however, a prominent feature that the mean residence time in any finite spatial domain is finite.

This ensures, for example, that the asymptotic response of the corresponding anomalous dynamics to time-periodic
perturbations is finite. However, these asymptotical responses in both models can also be different in 
some details, reflecting
different physics.
For example, it has been recently shown that the traditional CTRW subdiffusion based on infinite MRTs displays
asymptotically dying response to periodic fields \cite{DOR} which is at odds with most experimental data 
on anomalous dielectric response, and nevertheless agrees with some \cite{Allegrini}. 
Normally, the anomalous response exhibits aging behavior.  However, the response normally decays not to zero, 
but to some finite value. 
The response to periodic fields can serve to distinguish among different
underlying physical mechanisms.  A generalization of the presented here modified CTRW and FFPE 
model to time-dependent fields is not a trivial matter at 
all and it is left for a separate study.

Furthermore, for subdiffusion and transport restricted by the time scale $t<\tau_c$ and spatial scale $x<L\sim 
(\kappa_{\alpha}\tau_c^{\alpha})^{1/2}$ it is not easy to distinguish between two different models discussed here
which display a very similar transient behavior. Calculation of the velocity autocorrelation function
as e.g. in  Ref. \cite{Weber}  can clearly prove that the viscoelastic mechanism is at work, or statistical
analysis of single trajectories \cite{Magd} can help to make the distinction.
 The common feature is, however, that in both cases the transient transport
is \textit{faster} than the asymptotically normal one, and the transient subdiffusion is faster 
than asymptotically normal diffusion in the unbiased case. The results on mRNA molecules subdiffusion 
in bacterial cells \cite{Golding} seem consistent with this paradoxical conclusion \cite{FNL12}.
Moreover, the experimental results on subdiffusion of the colloidal gold particles having a typical
size of proteins with radius $R=2.5$ nm in cytoplasm of living cells \cite{Guigas} delivers a further clear 
confirmation. Indeed, in Ref. \cite{Guigas} experimental results were parameterized as 
\begin{equation}
\langle \delta x^2(t)\rangle =  2D\tau_D\left [\frac{t}{\tau_D} +\left(\frac{t}{\tau_s}\right)^\alpha  \right ]\;.
\end{equation}
Comparison with our model yields $\tau_c=[\Gamma(1+\alpha)\tau_D/\tau_s^\alpha]^{1/(1-\alpha)}$, $\kappa_1=D$ and
$\kappa_{\alpha}=D\tau_c^{1-\alpha}$. The experimental values of $\alpha$, 
$\tau_D$ and $\tau_s$ can be found for various cells
in \cite{Guigas}. The corresponding asymptotic normal diffusion coefficient is  $D=r_0^2/(4\tau_D)$, where $r_0=0.21\;\mu$m,
is the microscope aperture size \cite{Guigas}. From this we can obtain $\kappa_1$, $\kappa_{\alpha}$ and $\tau_{c}$,
see in the Table I.
\begin{table}\label{Table1} 
\caption{Parameters derived from experimental data}
\begin{ruledtabular}
\begin{tabular}{|p{1.5cm}|p{0.7cm}|p{0.55cm}| p{0.75cm}| p{0.9cm}|p{1.25cm}|p{1.3cm}|}
Cell & $\alpha$ & $\tau_D$, ms & $\tau_s$, ms & $\tau_c$, sec & $\kappa_1$, $\mu$m$^2$/sec & $\kappa_{\alpha}$, $\mu$m$^2$/sec$^\alpha$ \\
\hline
HeLa & 0.51 & 91 & 0.271 & 30.3 & 0.121 & 0.644 \\ 
THLE & 0.48 & 89 & 0.329 & 16.7 & 0.124 & 0.536 \\
HepG2 & 0.52 & 88 & 0.767 & 37.6 & 0.125 & 0.715 \\
HeLa \tiny
+ 500 mM sucrose & 0.66 & 92 & 0.954 & 2875 & 0.119 & 0.931 \\
\end{tabular}
\end{ruledtabular}
\end{table}
For gold beads of this size in water $\kappa_{1,\rm water}=90\mu$m$^2$/sec. Therefore, from the experimental values of 
$\kappa_1$ in cytoplasm of selected eukaryotic cells (possessing cytoskeleton) one can conclude that 
cytoplasm  is about $\kappa_{1, \rm water}/\kappa_1\sim 750$ times 
more viscous than water and about two times less viscous than glycerol. On the maximal time scale of experimental
data in Ref. \cite{Guigas}, which is about 1 sec, the particles subdiffuse much faster than one would expect from
the normal diffusion in such a viscous environment. This paradoxical and overlooked fact should be taken into
account in any discussion of the influence of transient subdiffusion on living processes as it appears that subdiffusion
can bring
only advantages, contrary to some common expectations \cite{FNL12}. 

The author is confident that the modified CTRW and FFPE theory
of subdiffusive transport considered in  this work presents a viable alternative to the traditional one.
 It shares more similarities with 
viscoelastic subdiffusion and should be taken seriously as an 
ample modeling
framework with attracting physical features and clear experimental relevance.  \\ 

Support of this work by the Deutsche Forschungsgemeinschaft, grant GO 2052/1-1
is gratefully acknowledged.

\end{document}